\documentclass{emulateapj}
\usepackage{fancyhdr}

\usepackage{natbib}
\usepackage{epsf}
\usepackage{float}

\begin{document}
\pagestyle{fancy}

\title{X-ray properties of K-selected galaxies at
  $0.5<\lowercase{z}<2.0$: Investigating trends with stellar mass,
  redshift and spectral type}
\slugcomment{Accepted to ApJ}
\author{Therese~M.~Jones\altaffilmark{1},
  Mariska~Kriek\altaffilmark{1},
  Pieter~G.~van~Dokkum\altaffilmark{2},
  Gabriel~Brammer\altaffilmark{3}, Marijn~Franx\altaffilmark{4},
  Jenny~E.~Greene\altaffilmark{5}, Ivo Labb{\'e}\altaffilmark{4},
  Katherine~E.~Whitaker\altaffilmark{2}}

\altaffiltext{1}{Department of Astronomy, University of California,
  Berkeley, CA 94720, USA {\it tjones@astro.berkeley.edu, mkriek@berkeley.edu}}
\altaffiltext{2}{Department of Astronomy, Yale University, New Haven,
  CT 06520-8101, USA}
\altaffiltext{3}{European Southern Observatory, Alonso de Córdova 3107, Casilla 19001, Vitacura, Santiago, Chile}
\altaffiltext{4}{Sterrewacht Leiden, Leiden University, NL-2300 RA
  Leiden, The Netherlands}
\altaffiltext{5}{Department of Astrophysical Sciences, Princeton University, Princeton, NJ 08544, USA}

\defcitealias{Hopkins11b}{b)}
\defcitealias{Hopkins11a}{Hopkins (2011a}
\defcitealias{Freeke11b}{b)}
\defcitealias{Freeke11a}{van de Voort 2011a}
\begin{abstract} 

We examine how the total X-ray luminosity correlates
with stellar mass, stellar population, and redshift for a $K$-band
limited sample of $\sim 3500$ galaxies at $0.5<z<2.0$ from the
NEWFIRM Medium Band Survey in the COSMOS field.  The galaxy sample is
divided into 32 different galaxy types, based on similarities between
the spectral energy distributions.  For each galaxy type, we further divide
the sample into bins of redshift and stellar mass, and perform an
X-ray stacking analysis using the Chandra COSMOS (C-COSMOS) data.  We
find that full band X-ray luminosity is primarily increasing with
stellar mass, and at similar mass and spectral type is 
higher at larger redshifts.  When comparing at the same stellar mass,
we find that the X-ray luminosity is slightly higher for younger
galaxies (i.e., weaker 4000\AA~breaks), but the scatter in this
relation is large. We compare the observed X-ray luminosities to those expected
from low and high mass X-ray binaries (XRBs). For blue galaxies, XRBs
can almost fully account for the observed emission, while for older galaxies
with larger 4000 \AA~breaks, active galactic nuclei (AGN) or hot gas dominate
the measured X-ray flux.  After correcting for XRBs, the X-ray
luminosity is still slightly higher in younger galaxies, 
although this correlation is not significant.  AGN appear to be a larger component of
galaxy X-ray luminosity at earlier times, as the hardness ratio increases with redshift.  Together with the slight increase
in X-ray luminosity this may indicate more obscured AGNs or higher accretion rates at earlier times.

\end{abstract}

\section{Introduction}
\label{sec:1}

Supermassive black hole growth and galaxy growth have long been thought
to be correlated, as supported by the Magorrian and
M$_{\mathrm{bh}}-\sigma$ relations \citep{Magorrian98, Ferrarese00,
 Gebhardt00}. Both active galactic nuclei (AGN) activity and star
formation peak at $z \sim 2$ (e.g., \citealt{Madau98,Merloni04}), and
the processes appear to be correlated in both AGN host and normal
galaxy samples (e.g., \citealt{Canalizo01,Kauffmann03}).  However, the
connection between star formation activity and black hole mass
accretion in galaxies is not well-understood.

Several studies suggest that AGN activity is needed to explain the
cutoff at the bright end of the galaxy luminosity function and the halting of star formation in the most
massive galaxies (\citealt{Croton06} and references therein).  This
theory is supported by the observation that AGN activity lags star formation by
$10^7-10^8$ years \citep{Davies07, Schawinski09, Wild10}, and that many AGNs
lie in the `green valley' (e.g., \citealt{Schawinski10, Xue10})-- the sparsely populated region between the red sequence and the blue cloud \citep{Faber07}.

 However, \citet{Rosario13b}
find that out to $z\sim2$ AGNs are preferentially hosted by 
star-forming, rather than quiescent or quenching galaxies.  Additionally, dust attenuation may redden the color of many AGN host galaxies (e.g., \citealt{Aird12},
\citealt{Cardamone10},\citealt{Brammer11}), implying that star
formation quenching by AGNs may not control the transition
between blue and red galaxies. 
Instead, the connection between AGN and star
formation activity may simply reflect the fact that both processes
require a large gas supply, with possible delays between star
formation and AGN activity caused by inefficient black hole fueling. 
For example, simulations by
\citetalias{Hopkins11a}, \citetalias{Hopkins11b} suggest that star
formation peaks as a function of gas density, while torques regulating
further inflow are relatively inefficient and limit black hole growth,
leaving a reservoir of gas long after the gas supply for star
formation is exhausted.  

While it is difficult to disentangle the contributions of star
formation and AGN activity of high redshift galaxies from optical
observations alone,  X-ray imaging acts as a better tracer of AGN
activity, with unobscured AGNs emitting heavily in both hard and soft
bands, and obscured AGN dominated by hard X-ray emission
(e.g. \citealt{Churazov02}).  Galactic X-ray emission also traces
stellar activity, through low mass X-ray binaries
(LMXBs), whose primarily soft X-ray emission correlates with stellar
mass, as well as high mass X-ray binaries (HMXBs) and supernova, whose
harder emission is reflective of star formation.  Soft X-ray emission
from hot gas is also apparent in the halos of massive ellipticals 
(e.g., \citealt{Grimm03,Gilfanov04a,Gilfanov04b, David06}).

\begin{figure}[t]
\centering
\epsfxsize=3.0in \epsfysize=3.0in
\epsfbox{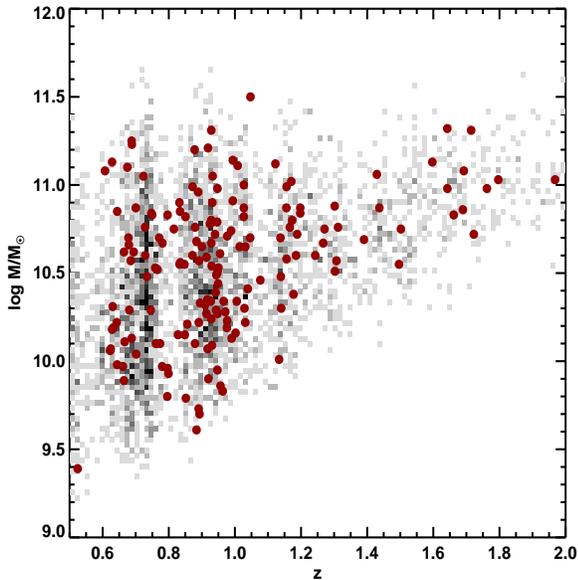}
\setlength{\belowcaptionskip}{30pt} 
\caption{Mass redshift distribution of the NMBS galaxy sample used in
 this study, with
  direct X-ray detections in red, and source density in gray scale.  X-ray exposures range from
  80-160ks, at a sampling of  $\sim 1/"$pixel. 
\label{fig:mz_dist}}
\end{figure}

\begin{figure}[t]
\centering
\epsscale{0.9}
\epsfxsize=3.0in \epsfysize=4.2in
\epsfbox{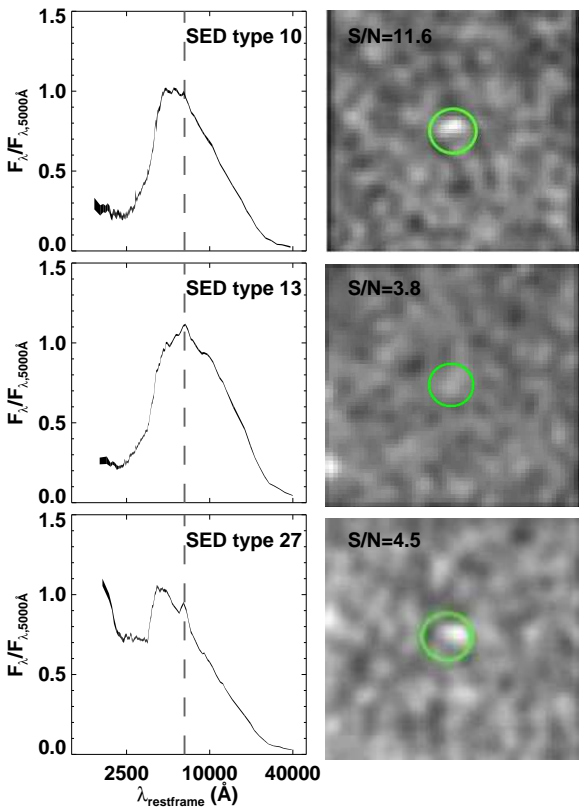}
\setlength{\belowcaptionskip}{10pt} 
\caption{Left: Three examples of composite SEDs (from Fig. 4 of
  \citealt{Kriek11}), with dashed line labeling $\rm{H \alpha}$,
  ordered by increasing $\rm{H \alpha}$ strength.  Right:
  Corresponding 30''x30'' X-ray stacks.  Green circles
  represent 5'' aperatures around the location of the NIR detection.
  External sources in this stack are not masked.
\label{fig:stacks}}
\end{figure}

The X-ray galaxy connection has been systematically studied at low
redshift, providing a rough census of X-ray luminosity with stellar mass,
galaxy type, and redshift
\citep[e.g.,][]{Symeonidis11,Lehmer07,Lehmer08}. For example,
\cite{Lehmer07} find that the AGN fraction for massive early type
galaxies increases with redshift, while the X-ray emission in less
massive galaxies evolves very little out to $z=0.7$. Late-type
galaxies evolve significantly in X-ray luminosity with redshift, and
furthermore their X-ray emission shows a strong dependence on stellar mass
\citep{Lehmer08}. At higher redshifts, X-ray galaxy studies
have primarily focused on direct X-ray detections (e.g.,
\cite{Silverman09,Kocevski12, Mullaney12,Santini12}) or on specific galaxy
populations, such as Lyman break galaxies (LBGs, Steidel et al. 1996),
submillimeter bright galaxies (SMGs), and distant red galaxies (DRGs,
Franx et al. 2003). For example, \cite{Laird10} find that the X-ray
emission is dominated by an AGN in only 15\%
of the SMGs out to $z=4$ in the Chandra Deep Field North \citep[see
also][]{Georg11}, 
while for LBGs and DRGs the X-ray emission seems almost completely accounted for
by star formation (e.g., Reddy \& Steidel 2004; Rubin et
al. 2004). However, these different selection techniques target galaxies at
different redshifts and stellar masses, and even combined are not
representative of the full distant galaxy population. Hence, a complete census
of the X-ray properties of distant galaxies is still missing.

Using the high-quality near-infrared photometry from the NEWFIRM Medium Band Survey
(NMBS,\citealt{vD09, Whitaker11}) in combination with Chandra data, it is possible
to systemically study the X-ray properties of an observed
  K-band complete sample of
distant galaxies.  In this study, we utilize the $\sim 3500$ galaxy
from the NMBS and 32 SED templates of \cite{Kriek11}, along with the
overlapping C-COSMOS field, to better characterize the star
formation-AGN connection out to $z \sim 2$.  With accurate
$W_{\mathrm{H}\alpha}$ and $D(4000)$ measurements from \cite{Kriek11},
indicating instantaneous star formation rates and stellar population ages,
along with separate hard and soft band X-ray luminosities, we attempt
to correlate AGN activity with star formation properties, stellar
mass, and redshift.

\begin{deluxetable*}{cccccccc}
\tablecaption{Stellar population properties and X-ray luminosities}
\tablenum{1}
\tablehead{\colhead{SED type} & \colhead{$N$ \tablenotemark{a}}  &  \colhead{$D(4000)$} &
  \colhead{$W{\mathrm{H\alpha}}$ }
  & \colhead{log($SSFR$)} &
  \colhead{$A_V$} & \colhead{log${L_x}$\tablenotemark{b}} &
  \colhead{log$\frac{L_x}{M/10^{10}M{\odot}}$\tablenotemark{c}} \\ &  & & (\AA)
  &  [yr$^{-1}$] &  (mag) & [erg/s] & [erg/s] }

\startdata
1 & 121 & 1.95 & 14.4 & -11.79 & 0.4 & $ 41.4 ^{+ 0.1 }_{- 0.2} $ & $ 40.4 ^{+ 0.1 }_{- 0.2} $ \\ [1.5ex]
2 & 291 & 1.94 & 14.4 & -11.79 & 0.2 & $ 41.6 ^{+ 0.2 }_{- 0.3} $ & $ 40.6 ^{+ 0.2 }_{- 0.3} $ \\ [1.5ex]
3 & 436 & 1.81 & 6.2 & -24.93 & 0.5 & $ 40.5 ^{+ 0.2 }_{- 0.3} $ & $ 39.7 ^{+ 0.2 }_{- 0.3} $ \\ [1.5ex]
4 & 211 & 1.77 & 2.1 & -18.35 & 0.3 & $ 41.4 ^{+ 0.2 }_{- 0.2} $ & $ 40.7 ^{+ 0.2 }_{- 0.2} $ \\ [1.5ex]
5 & 95 & 1.70 & 24.6 & -10.59 & 0.7 & $ 41.8 ^{+ 0.1 }_{- 0.2} $ & $ 41.0 ^{+ 0.1 }_{- 0.2} $ \\ [1.5ex]
6 & 22 & 1.67 & 19.1 & -11.67 & 0.1 & $ 41.7 ^{+ 0.2 }_{- 0.3} $ & $ 40.8 ^{+ 0.2 }_{- 0.3} $ \\ [1.5ex]
7 & 72 & 1.66 & 20.5 & -11.29 & 0.2 & $ 41.9 ^{+ 0.1 }_{- 0.2} $ & $ 41.0 ^{+ 0.1}_{- 0.2} $ \\ [1.5ex]
8 & 64 & 1.66 & 38.7 & -10.49 & 1.2 & $ 41.6 ^{+ 0.1 }_{- 0.2} $ & $ 40.8 ^{+ 0.1}_{- 0.2} $ \\ [1.5ex]
9 & 94 & 1.65 & 42.7 & -10.59 & 1.1 & $ 41.3 ^{+ 0.2 }_{- 0.3} $ & $ 40.5 ^{+ 0.2 }_{- 0.3} $ \\ [1.5ex]
10 & 60 & 1.62 & 33.1 & -10.58 & 0.5 & $ 42.2 ^{+ 0.2 }_{- 0.2} $ & $ 41.4 ^{+ 0.2 }_{- 0.2} $ \\ [1.5ex]
11 & 95 & 1.58 & 39.1 & -10.4 & 0.7 & $ 41.9 ^{+ 0.1 }_{- 0.2} $ & $ 41.1 ^{+ 0.1 }_{- 0.2} $ \\ [1.5ex]
12 & 46 & 1.58 & 17.7 & -10.58 & 1.4 & $ 41.9 ^{+ 0.2 }_{-0.3} $ & $ 41.0 ^{+ 0.2 }_{-0.3} $ \\ [1.5ex]
13 & 114 & 1.45 & 63.3 & -10.71 & 2.4 & $ 41.8 ^{+ 0.2 }_{- 0.3} $ & $ 41.0 ^{+ 0.2 }_{- 0.3} $ \\ [1.5ex]
14 & 71 & 1.45 & 69.4 & -9.66 & 2.6 & $ 41.5 ^{+ 0.1 }_{- 0.2} $ & $ 40.8 ^{+ 0.1 }_{- 0.2} $ \\ [1.5ex]
15 & 64 & 1.40 & 61.7 & -10.34 & 2.8 & $ 41.4 ^{+ 0.2 }_{- 0.2} $ & $ 40.7 ^{+ 0.2 }_{- 0.2} $ \\ [1.5ex]
16 & 155 & 1.40 & 67.4 & -9.92 & 2.1 & $ 41.5 ^{+ 0.1 }_{- 0.2} $ & $ 40.9 ^{+ 0.1 }_{- 0.2} $ \\ [1.5ex]
17 & 103 & 1.40 & 66.2 & -9.66 & 1.9 & $ 41.8 ^{+ 0.1 }_{- 0.2} $ & $ 40.2 ^{+ 0.1 }_{-0.2} $ \\ [1.5ex]
18 & 95 & 1.38 & 86.0 & -9.34 & 2.3 & $ 41.6 ^{+ 0.1 }_{- 0.2} $ & $ 41.0 ^{+ 0.1}_{-0.2} $ \\ [1.5ex]
19 & 33 & 1.37 & 86.1 & -9.66 & 2.6 & $ 41.6 ^{+ 0.1 }_{- 0.2} $ & $ 40.9 ^{+ 0.1 }_{-0.2} $ \\ [1.5ex]
20 & 34 & 1.37 & 108.0 & -8.81 & 2.6 & $ 41.8 ^{+ 0.3 }_{- 0.8} $ & $ 41.1 ^{+ 0.3}_{- 0.8} $ \\ [1.5ex]
21 & 139 & 1.35 & 79.5 & -9.15 & 1.9 & $ 41.6 ^{+ 0.1 }_{- 0.2} $ & $ 41.1 ^{+ 0.1 }_{- 0.2} $ \\ [1.5ex]
22 & 82 & 1.34 & 66.9 & -10.24 & 1.7 & $ 41.5 ^{+ 0.2 }_{-0.4} $ & $ 41.1 ^{+ 0.2 }_{- 0.4} $ \\ [1.5ex]
23 & 51 & 1.33 & 92.4 & -8.69 & 2.2 & $ 41.2 ^{+ 0.3 }_{-0.9} $ & $ 40.7 ^{+ 0.3 }_{-0.9} $ \\ [1.5ex]
24 & 26 & 1.30 & 130.7 & -9.25 & 2.1 & $ 42.1 ^{+ 0.2 }_{- 0.4} $ & $ 41.5 ^{+ 0.2 }_{-0.4} $ \\ [1.5ex]
25 & 159 & 1.30 & 94.1 & -10.71 & 1.4 & $ 41.1 ^{+ 0.2 }_{-0.4} $ & $ 40.8 ^{+ 0.2 }_{- 0.4} $ \\ [1.5ex]
26 & 80 & 1.30 & 107.1 & -8.81 & 1.9 & $ 41.5 ^{+ 0.2 }_{-0.3} $ & $ 41.1 ^{+ 0.1 }_{- 0.3} $ \\ [1.5ex]
27 & 108 & 1.27 & 105.8 & -8.78 & 1.6 & $ 41.8 ^{+ 0.2}_{- 0.2} $ & $ 41.6 ^{+ 0.2 }_{- 0.2} $ \\ [1.5ex]
28 & 144 & 1.25 & 106.2 & -9.92 & 1.2 & $ 41.8 ^{+ 0.2 }_{- 0.3} $ & $ 41.7 ^{+ 0.2 }_{- 0.3} $ \\ [1.5ex]
29 & 64 & 1.22 & 102.8 & -8.61 & 1.4 & $ 41.7 ^{+ 0.1 }_{- 0.2} $ & $ 41.6 ^{+ 0.1}_{- 0.2} $ \\ [1.5ex]
30 & 142 & 1.21 & 104.0 & -8.65 & 1.2 & $ 41.7 ^{+ 0.2 }_{- 0.4} $ & $ 41.7 ^{+ 0.3 }_{-0.4} $ \\ [1.5ex]
31 & 101 & 1.18 & 109.7 & -8.65 & 1.0 & $ 41.6 ^{+ 0.3 }_{- 0.7} $ & $ 41.7 ^{+ 0.3 }_{-0.7} $ \\ [1.5ex]
32 & 50 & 1.13 & 138.5 & -8.51 & 0.9 & $ 42.0 ^{+ 0.3 }_{-1.0} $ & $ 42.1 ^{+ 0.3 }_{- 1.0} $ \\ [1.5ex]

\enddata

\footnotetext[1]{Number of
      galaxies per spectral type}
\footnotetext[2]{Average full
      band X-ray luminosity of the stack}
\footnotetext[3]{Average full
    band X-ray luminosity weighted by mass of the stack}

\tablecomments{The 4000\AA~break ($D(4000)$) strength, H$\alpha$
  equivalent width  (WH$\alpha$), specific SFRs, dust attenuation ($A_v$) are adopted
  from Kriek et al. The specific SFR (SSFR) and $A_V$s are derived by
  comparing the composited SEDs to the \citet{BC03}
  stellar population synthesis models using a \citet{Calzetti00} law, a
 \citet{Salpeter55} IMF, and a delayed exponential star formation history. The
  4000\AA~breaks and H$\alpha$ equivalent widths are directly measured from the composite SEDs}
\label{tab:sed_props}
\end{deluxetable*}

\section{Data and Methods}
\label{sec:2}

In this work, we make use of the photometric catalogs from the NMBS in the COSMOS
field \citep{vD09, Whitaker11}.  The NMBS uses five custom NIR
filters (three J-band and two H-band), covers 0.2 deg$^2$ of the COSMOS field
\citep{Scoville07}, and is complemented by FUV-MIR data sets
\citep[e.g.,][]{Sanders07,Capak07,Barmby08,Erben09,Hildebrandt09}.
The photometric redshifts and stellar population properties
are derived using EAZY \citep{Brammer08} and FAST \citep{Kriek09}.

We utilize the NMBS sample of $\sim3500$ galaxies at $0.5<z<2.0$ of 
\citealt{Kriek11} (with mass-redshift distribution displayed in
Fig.~\ref{fig:mz_dist}), and employ the spectral
classification therein. In summary, the galaxies were divided into 32 subsamples
based on similarities in their spectral energy distributions
(SEDs). First, the rest-frame photometric SEDs of all individual
galaxies were compared, and based on the similarities in their full
SED shape, it was decided whether or not two galaxies were
analogs. Next, the galaxy with most analogs was identified, which
together formed the first subsample. After removal of all galaxies of
the first subsample, the next galaxy with most analogs was
identified. This procedure was repeated until nearly all galaxies were
divided. Eventually, some galaxies were re-assigned, if they would
have been a better match with a later iteration.  Thus, our
classification method is model independent.

\begin{figure*}[ht]
\centering
\epsscale{0.9}
\epsfxsize=6.5in \epsfysize=3.25in
\epsfbox{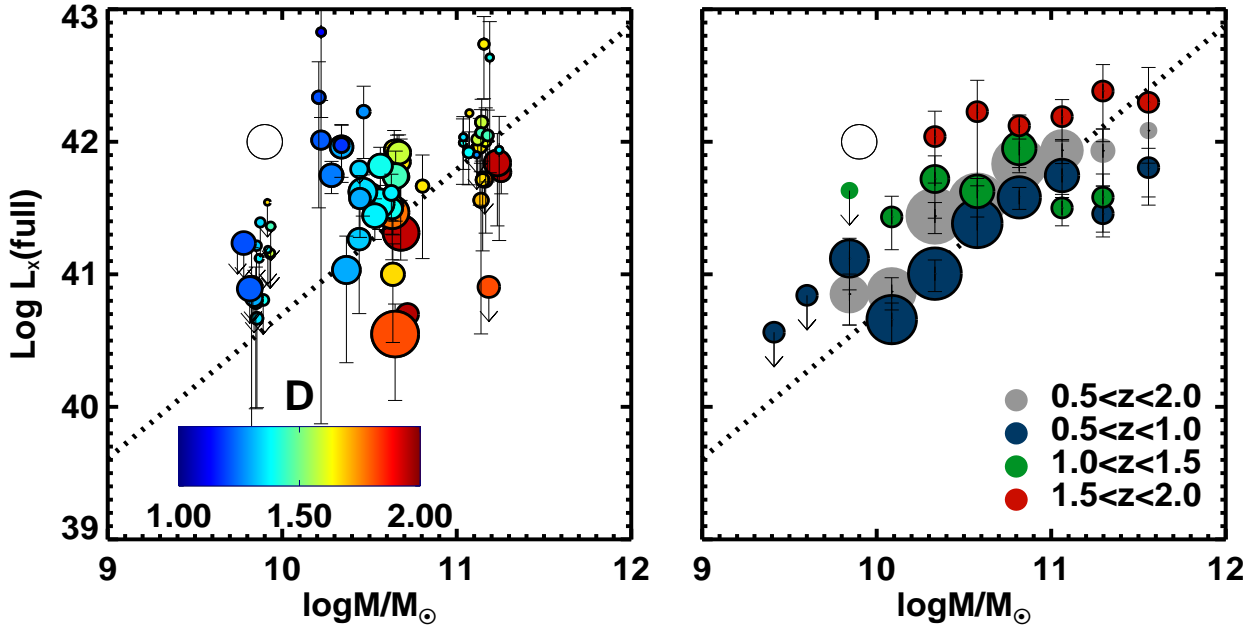}
\caption{Average full band (0.5-7 keV) X-ray luminosity vs. mass for each
  composite SED mass bin, color-coded by $D(4000)$
  and point size indicating the number of galaxies per bin, with $1\sigma$
  limits (left). The white circle represents
  galaxies not fit by any of the 32 SED templates.  We additionally plot all composite SEDs binned by mass,
  rather than spectral type (right).  Gray points indicate the entire
  sample, while colored points show the sample subdivided into three redshift bins.  There appears to be an
  approximately linear correlation, with a slope of 1 (dashed line),
  with higher luminosities observed at higher masses and redshifts.
\label{fig:luminm}}
\end{figure*}

\begin{figure}[t]
\centering
\vspace{0.2in}
\epsfxsize=3.0in \epsfysize=3.0in
\epsfbox{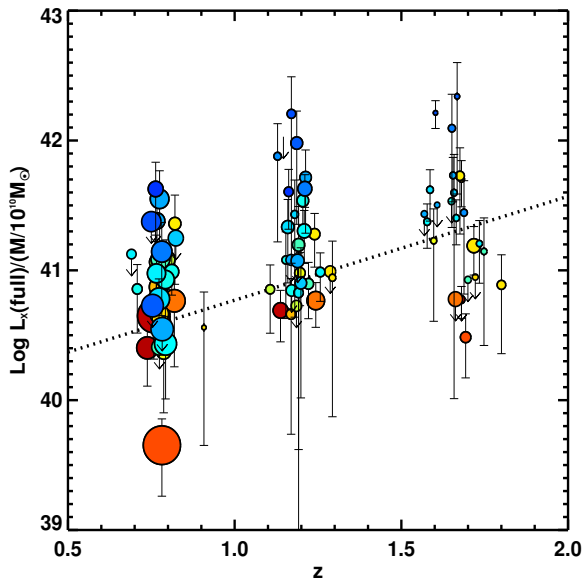}
\setlength{\belowcaptionskip}{30pt} 
\caption{Average full band (0.5-7 keV) X-ray luminosity vs. z for each
  composite SED bin, normalized by mass, and color-coded by $D(4000)$
  (see Fig~\ref{fig:luminm}).  Points are drawn by bin size, with $1\sigma$
  limits.  There appears to be at best a slight
  correlation between luminosity and redshift once the effects of mass
  evolution are removed.
\label{fig:lz}}
\end{figure}

\begin{figure}[t]
\centering
\vspace{0.2in}
\epsfxsize=3.0in \epsfysize=3.0in
\epsfbox{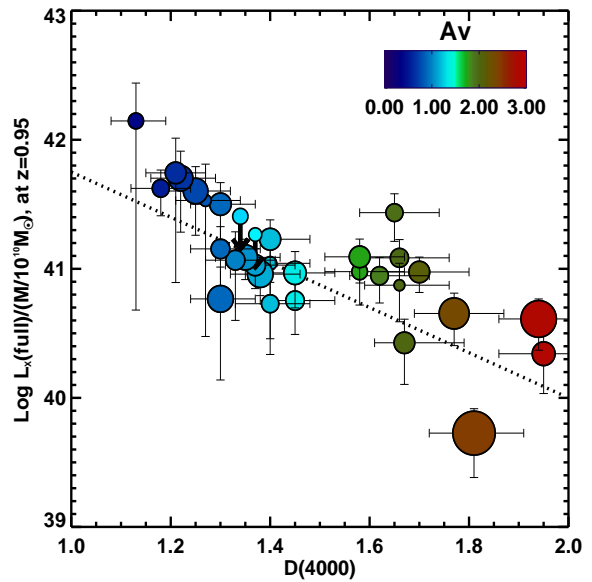}
\caption{Full band (0.5-7 keV) X-ray luminosity vs. $D(4000)$,
   normalized by mass and
  corrected to the average redshift of the sample using the best fit
  $L_x-z$ relation, and color-coded by Av (points drawn by bin size, with $1\sigma$
  limits).  X-ray luminosity per unit mass appears to be
  anti-correlated with $D(4000)$, with a difference of approximately a
  dex between low and high values of $D(4000)$.  This may suggest
  higher black hole accretion rates in blue galaxies, or significant
  X-ray contribution from star formation.
\label{fig:ldav}}
\end{figure}

A composite SED is constructed for each subsample by
  combining the rest-frame photometry of all galaxies in the bin.   Each subsample contains 22-436 galaxies for which there is X-ray
 coverage as well. 
The final composite SED sample spans almost the entire
galaxy population at these redshifts.  Each composite SED
shows detailed spectral features such as the H$\alpha$+[N{\sc II}]+[S{\sc II}]
and H$\beta$+[O{\sc III}] emission lines, Balmer or 4000\AA~breaks,
MgII absorption at 2800\AA, the continuum break at 2640\AA, and the
dust absorption feature at 2175\AA ~(see Fig. 4 of
\citealt{Kriek11}, and three examples in Fig. ~\ref{fig:stacks}).  As
the sample is K-band (and not stellar mass) 
selected, it will be slightly biased toward star-forming galaxies with
lower $M/L$ ratios. Furthermore, as our
 photometric redshift code only uses stellar population templates, we have a bias against AGN-dominated SEDs.

Although $D(4000)$ measures stellar population age (as stellar
opacity increases with age and strengthens absorption
lines of ionized metals such as Ca$\mathrm{II}$ H and K), rather than
ongoing star formation activity,
\cite{Kriek11} show a clear correlation between $D(4000)$,
$W_{\mathrm{H}\alpha}$, and the specific star formation rate (SSFR).  
Galaxies with shallow $4000 \AA$ breaks have high
$W_{\mathrm{H}\alpha}$ and are fitted by stellar population models
with high SSFRs, while galaxies with strong breaks have low values of
$W_{\mathrm{H}\alpha}$ and low SSFRs.  Basic properties of each SED
type are shown in Table~\ref{tab:sed_props}.

In this paper, we study the X-ray properties of the different SED
types.  We make use of the C-COSMOS survey \citep{Elvis09,Puccetti09},
which has an effective exposure of 80-160ks, and a sampling of
1''/pixel.  Merged images of 0.5-2 keV (soft), 2-7 keV (hard), and
0.5-7 keV (full) are publicly available, along with the point source
catalog of \cite{Elvis09}.  Each field of the C-COSMOS survey is
observed with up to six overlapping pointings, ensuring relatively
uniform sensitivity, despite differing point-spread functions
(\citealt{Puccetti09}).

We obtain 60''x60'' C-COSMOS cutouts of each galaxy field from
\cite{Kriek11}.   As a vast majority of the galaxies are not detected
in the X-ray, we opt to stack by SED type, and further subdivide by
redshift and/or stellar mass.  We note that direct
detections are included in the stacks and thus the further analysis.  We select source counts
within a 5'' radius of the NIR detection, and make background estimates by
taking 30 random 5'' radius apertures in each frame, excluding 5''
apertures around the optical source or known X-ray sources
\citep{Elvis09}.

We then de-redshift the source according to its photometric redshift (or spectroscopic
redshift when available), using a photon index of $\Gamma = 1.1$, to
determine a source luminosity, via the equation
$$L_{E_1-E_2}=4\pi d_L^2f_{E_1-E_2}(1+z)^{\Gamma -2},$$ where
$f_{E_1-E_2}$ is the observed frame emission in the $E_1-E_2$
bandpass, and $d_L$ is the luminosity distance.  We explore alternate photon indices,
as the steepness of galaxy X-ray spectra varies according to the
strength of the various types of X-ray sources, and find that the
observed trends in X-ray luminosity do not change significantly with
differing $\Gamma$  (see \ref{sec:3.2} for further discussion).  We
then determine the average luminosity of all sources per bin.  

To determine the error in each stack, we bootstrap resample the
sources, subtracting a random background aperture (of the selected
30) from each frame.  The error is the standard
deviation of the average stack luminosities of the bootstrap simulations.  Stacked
detections of less than $3\sigma$ (as determined by the bootstrapped
background estimates) were deemed insignificant.  Sample
stacks for three SED types of 
full-band X-ray data are shown in
Fig.~\ref{fig:stacks}


\section{Analysis}
\label{sec:3}
In this section, we explore X-ray properties of the galaxy population
at $0.5<z<2.0$.  To determine the dependence of X-ray luminosity on
galaxy type, we seek to disentangle the effects of redshift and
stellar mass.  We attempt to remove the effects of mass and redshift
by splitting galaxies into bins, and eventually normalizing by stellar
mass (\S\ref{sec:3.1}).  In \S\ref{sec:3.2}, we look at
X-ray hardness as well as expected XRB contribution to galaxy
luminosity in an attempt to distinguish the sources of X-ray flux.

\subsection{X-ray correlations}
\label{sec:3.1}

As each SED type probes a different mass distribution, we first
explore the X-ray luminosity as a function of stellar mass by
splitting each sub sample into three mass bins ($\log M/M_{\odot} < 10, 10<\log M/M_{\odot} < 11$,
and $\log M/M_{\odot} >11$)  in the left panel of Fig.~\ref{fig:luminm}. The X-ray
luminosity is correlated with stellar mass, 
such that more massive galaxies emit at higher rates regardless of
stellar age.  We determine a least squares fit to the
data via resampling, assuming a normal distribution
  with standard deviation as calculated by the bootstrap errors for
  each data point.  We calculate the line of best fit for each
  resampling, weighing individual points by the number of galaxies in
  each stack, and determine the average and standard deviations of the
  fits to be $\log L_x \sim 1.1
(\pm 0.1) 
\log M +30 (\pm 1)$.

To further examine this relation we stack all galaxies in 10 bins  of
stellar mass, and show these as the large gray filled circles in the
right panel of right panel of Fig.~\ref{fig:luminm}, with errors calculated by resampling the
galaxies in each bin. We further divide these 10 mass bins into 3
redshift bins (colored), and find that higher redshift galaxies have
higher X-ray luminosities at the same mass, when compared to lower
redshift galaxies. At $z<1$ there is a clear correlation between the X-ray luminosity and stellar mass, which is broadly consistent with a one-to-one relation. The X-ray luminosity is increasing with redshift, and the trend with mass seems to flatten. As we probe different mass ranges at different redshifts, the increase of X-ray luminosity with redshift further steepens the correlation between X-ray luminosity and stellar mass. Thus,
to further examine the trend with redshift and other galaxy
properties, we normalize the X-ray luminosity by the stellar mass from
here on.

As $L_x/M$ is roughly constant, we may
estimate the typical accretion rate of the galaxy sample.  Assuming a $10^{7}
\mathrm{M_{\odot}}$ black hole 
in a galaxy of stellar mass $10^{10} \mathrm{M_{\odot}}$, and a
bolometric correction of $L_{Bol} \sim 15.8 L_x(2-10 \mathrm{keV})$ \citep{Ho09}, we
find the average Eddington ratio of our sample is $L_{Bol}/L_{Edd}
\sim 10^{-3}$.  We will argue below that accretion is the dominant
source of emission.

We further split the sample by redshift ($z<1.0, 1.0<z<1.5,$
and $z>1.5$; for 9 total bins) and restack using the method described
in Section~\ref{sec:2}, normalizing by stellar mass (such that we
measure total($L_{x})/$total(stellar mass).  Fig.~\ref{fig:lz} shows  $\log L_{x}$(full) / $M$ vs. $z$ for
all bins combined.   We find $\log L_x/(M/10^{10}M_{\odot})  \sim 0.8 (\pm 0.2) z+40.0 (\pm
0.2)$, suggesting that X-ray luminosity
is higher at earlier times for galaxies of similar mass.  This trend does not change upon making a minimum mass cut
  at log$M/M_{\odot} \sim 10.6$, the minimum mass detectable at $z
  \sim 2$.  

To remove the effects of any redshift
correlation, we normalize by the above luminosity-redshift relation
(as if all galaxies were at the average redshift, z=0.95)
after recombining all mass/redshift bins, 
and divide total stack luminosity
by total stellar mass contained in each stack, to explore trends in
$D(4000)$ independent of galaxy mass and redshift.  Fig.~\ref{fig:ldav} shows
mass and redshift-normalized X-ray luminosity as a function of stellar age (as
measured by the strength of the 4000 \AA~break), with the colors indicating
$\mathrm{A_v}$.  We find the following correlation between the
normalized X-ray luminosity as a function of stellar population age: $\log L_x/(M/(10^{10}M_{\odot})) \sim
-1.8 (\pm 0.4)D(4000) + 43.5 (\pm 0.6)$, suggesting that there is
a moderate correlation between $D(4000)$ and galaxy X-ray luminosity.  This correlation may have caused the flattening or increase of the X-ray luminosity for the lowest mass bins for each redshift interval in Figure 3b, as these bins are dominated by star-forming galaxies with slightly higher X-ray luminosities. It may also explain the flattening at the highest masses, as quiescent galaxies with slightly lower X-ray luminosities are dominating these bins.

\begin{figure}[t]
\centering
\epsfxsize=3.0in \epsfysize=3.0in
\epsfbox{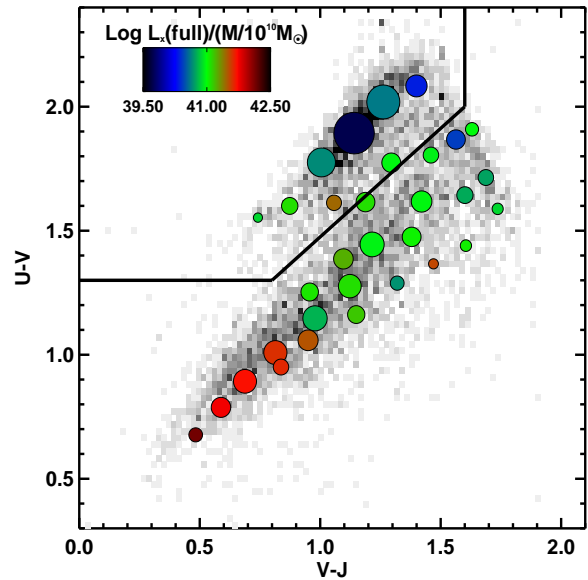}
\setlength{\belowcaptionskip}{30pt} 
\caption{Rest frame U-V vs. V-J color for each SED type, color-coded
  by full band (0.5-7keV) normalized
  X-ray luminosity, corrected to the average redshift of the
    sample ($z=0.95$) using the best fit
  $L_x-z$ relation, and scaled by bin size.   The full NMBS sample
  with K-band S/N $>25$ is shown in grey.  Passive red galaxies (top
  left; bounded by the \cite{Williams09} division separating
    quiescent and SFGs.) have an average luminosity per stellar mass almost a dex lower
  than that of blue galaxies.
\label{fig:uvvj}}
\end{figure}

We also explore the X-ray properties of the 32
different galaxy types of Fig.~\ref{fig:ldav} in U-V and V-J color space in Fig.~\ref{fig:uvvj}, in
comparison to the color of all detected NMBS galaxies with $S/N > 25$ (grey scale).
This diagram is a powerful diagnostic in distinguishing the blue and
red galaxy populations, and separates the red
passively-evolving galaxies from the dusty red star-formers; quiescent
galaxies are primarily located in the upper left quadrant of this
diagram \citep{Labbe05, Wuyts07,Williams09}.  The X-ray luminosity
decreases along the star forming sequence, and the quiescent
galaxies have an average X-ray luminosity per unit mass almost a dex lower than that
of the bluest galaxies in the sample.

To first order, we observe a linear trend in galaxy X-ray luminosity
as a function of stellar mass (Fig.~\ref{fig:luminm}).  After normalizing by stellar mass, we
observe underlying redshift and stellar population age relations,
finding that galaxy X-ray luminosity increases at higher redshifts
(Fig.~\ref{fig:lz}), and
also is lower for older stellar populations (Fig.~\ref{fig:ldav}).

\subsection{Origin of the X-ray flux}
\label{sec:3.2}

\begin{figure*}[t]
\centering
\epsfxsize=7in \epsfysize=2.3in
\epsfbox{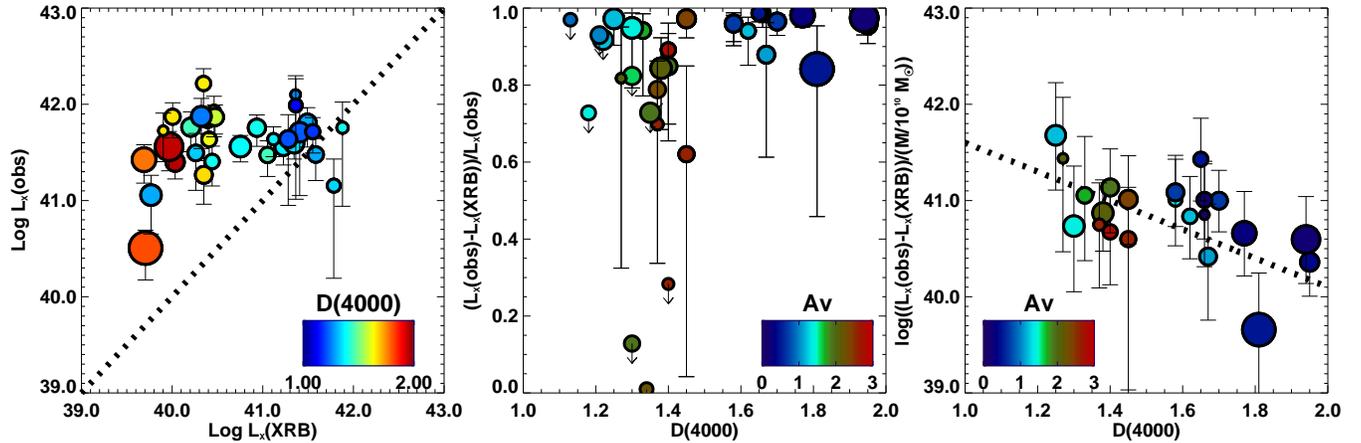}
\setlength{\belowcaptionskip}{30pt} 
\caption{Left: Average full band (0.5-7 keV) X-ray luminosity vs. expected
  X-ray binary (XRB) luminosity.   XRB luminosity is derived from NIR star formation rate/mass
  estimates of \cite{Kriek11} for each
  composite SED bin, color-coded by $D(4000)$.   Center:
The fraction of the
  luminosity not accounted for by XRBs vs. $D(4000)$, color-coded by
  $A_v$.  Right: Luminosity in excess of that predicted by XRBs, normalized
  to $10^{10}M_{\odot}$.   $1\sigma$ limits are labeled with arrows, and galaxies
  whose luminosity is over-predicted by the $Lx$-XRB relations are
  shown at an excess luminosity fraction of 0.  Luminosity in
  excess of that predicted by XRBs appears to be anti-correlated with
  $D(4000)$.
\label{fig:lx_expected1}}
\end{figure*}

In this section, we attempt to distinguish between the stellar, AGN, and gaseous
components of the observed X-ray flux.  We start by comparing the measured X-ray
flux with the expected X-ray flux from XRBs.  To determine the expected
X-ray flux from XRBs, we utilize the star formation rates
and stellar masses derived in \cite{Kriek11}, and X-ray luminosity
relations for high-mass X-ray binaries (HMXBs) and low-mass X-ray
binaries (LMXBs).  Luminosity, stellar mass, and star
formation rate relations are taken from \cite{Grimm03} and \cite{Gilfanov04a}, such that
$L_{2-10\mathrm{keV}}=6.7 \times 10^{39} \mathrm{erg s^{-1} SFR
  (M_{\odot}/year)}$ and $L_{x} (>10^{37} \mathrm{erg s^{-1}}) = 8.0
\times 10^{39} \mathrm{erg
  s^{-1}}/(M_{stellar}/10^{11}\mathrm{M_{\odot}})$ (see
\cite{Gilfanov04b}, and \cite{PR07} for further discussion of these relations). We plot the observed X-ray
luminosity vs. expected X-ray luminosity from XRBs in the left panel of
Figure~\ref{fig:lx_expected1}. The blue galaxies are
well-described by the X-ray luminosity of XRBs alone.  However, for
galaxies with strong 4000\,\AA\ breaks, the contribution from XRBs
cannot account for the full X-ray flux.  This implies that a large fraction of their X-ray
luminosities comes from AGN and/or hot gas contributions.  In the middle panel of
Figure~\ref{fig:lx_expected1} we show the relative contribution of
X-ray flux from non-stellar sources.  This plot illustrates
that the non-stellar X-ray sources contribute a significant fraction
of the X-ray flux, in particular for galaxies with larger 4000 \AA\
breaks (luminosities of several galaxies with smaller 4000 \AA\ breaks are
upper limits).  In the right panel of Figure~\ref{fig:lx_expected1} we show
the measured X-ray flux corrected for XRBs, normalized by stellar
mass, as a function of $D(4000)$.  Interestingly, we find that 
$\log L_x/M \sim -1.5 (\pm 0.1) D(4000) + 43.1 (\pm 0.2)$.
This suggests that the X-ray luminosity difference between red
  and blue galaxies is almost entirely driven by non-stellar sources. We note that galaxies with higher
values of $A_v$ may be dominated by emission from star formation
(middle and right panels); however, these measurements are not well-constrained.

\begin{figure}
\epsfxsize=3.0in \epsfysize=3.0in
\epsfbox{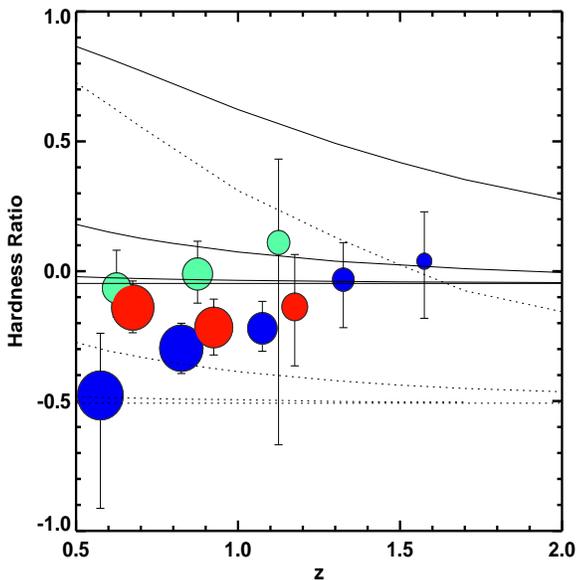}
\setlength{\belowcaptionskip}{30pt} 
\caption{Hardness ratios vs. redshift, color-coded by $D(4000)$, with
  $D(4000)<1.4$ in blue, $1.7>D(4000)>1.4$ in green, and $D(4000)>1.7$
  in red.  The sample is divided
  into 6 redshift bins, with points corresponding to bin size, and $1\sigma$ upper and lower limits.
 Lines indicate the column density of the obscuring medium
  that would produce the given hardness ratios, where solid indicates
  $\Gamma=1.1$ and dashed $\Gamma=2.1$ for obscuring column
  densities of log $N = 20, 21, 22, 23$~[cm$^-2$] (bottom-top).  While
  there is no clear correlation with hardness and $D(4000)$, the
  galaxies appear to be more obscured at higher redshift.
\label{fig:lx_expected}}
\end{figure}

\begin{figure}
\epsfxsize=3.0in \epsfysize=3.0in
\epsfbox{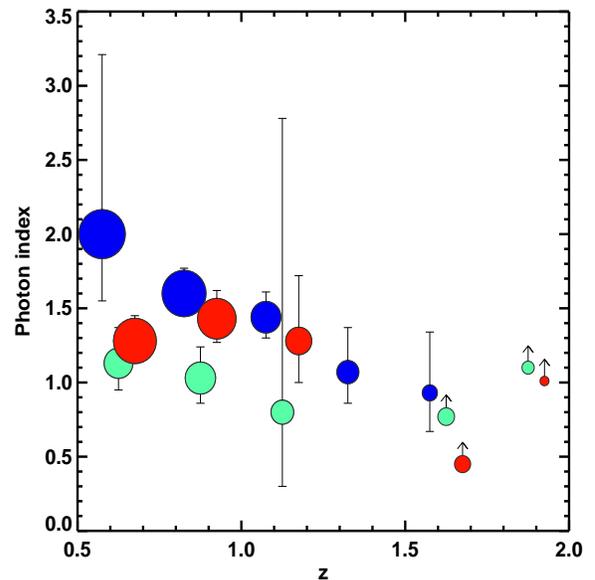}
\setlength{\belowcaptionskip}{30pt} 
\caption{Photon index as a function of redshift, as measured in 3
  different $D(4000)$ bins  color-coded by $D(4000)$, with
  $D(4000)<1.4$ in blue, $1.7>D(4000)>1.4$ in green, and $D(4000)>1.7$
  in red.  Photon index appears to decrease as
  redshift increases, suggesting galaxies are more obscured at higher redshifts.
\label{fig:hrz}}
\end{figure}

Next we explore the hardness ratio, given by $\mathrm{(H-S)/(H+S)}$,
where H is the number of hard band (2-8 keV) counts and S is the
number of soft band (0.5-2 keV) counts (Fig.~\ref{fig:lx_expected}),
as a function of $D(4000)$ and redshift to distinguish
probable X-ray sources.  We plot lines to determine the
hardness ratio that would be observed in a galaxy with the listed
obscuring column densities, assuming a photon index of 1.1 (solid) and
2.1 (dashed).  We similarly plot the photon index as a function of redshift by
combining the galaxies into $D(4000)$ and $z$ bins, and
measuring the hard and soft counts, assuming a constant hardness ratio (Fig.~\ref{fig:hrz}).   We find
that the obscuring column density increases as a function of redshift,
and may be higher for galaxies with intermediate values of $D(4000)$
(green).  Interestingly, these are galaxies with high values of $A_v$.  This suggests
a greater AGN contribution to the average galaxy X-ray luminosity at
higher redshifts, consistent with the observed increase in total X-ray
luminosity at the same redshifts.

\section{Discussion}

After correcting for an increase in X-ray luminosity with both stellar mass and redshift, we find a difference of 1 dex between the X-ray luminosities of the highest and lowest SFR galaxies.To separate the effects of black hole accretion and stellar luminosity
components, we subtract the expected X-ray binary
contribution and normalize by stellar mass, finding that a
significant fraction of the X-ray 
luminosity is from non-stellar sources.  This excess is highest
for passive red galaxies.  We note that both hot gas
and black hole accretion may contribute to differences in
$L_x$.

While hot gas likely remains an important
part of the X-ray luminosity function out to at least $z=1.4$
\citep{Tremmel13}, we note that its contribution  to the rest-frame soft
band decreases significantly at higher redshifts, as galactic hot gas
primarily emits at $kT \lesssim 1.0$ keV.
Additionally, X-ray emission from hot gas is unlikely directly correlated with star formation rate for
galaxies of similar masses.  The X-ray luminosities of local galaxies can vary by orders of magnitude
for galaxies of similar stellar or black hole masses
\citep{Fabbiano89, Pellegrini98, Sarazin01, Pellegrini10}.  The
cosmological hot gas accretion rate onto halos is predicted to plateau
at $z<1$, however the used models are still highly uncertain (\citetalias{Freeke11a},
\citetalias{Freeke11b}).  

We thus attribute the trends in X-ray luminosity to differences in AGN
accretion rates.  Previous studies have found similar trends in galaxy mass and
accretion rates, though
we probe lower black hole accretion rates, averaging $L_{Bol}/L_{Edd}
\sim 10^{-3}$ (assuming the $M_{BH}-\sigma$ relation, \citet{Magorrian98}). Both \cite{Bongiorno12}
and  \cite{Aird12} find an increase in AGN luminosity with stellar
mass out to $z=3$ and $z=1$, respectively, for objects several orders
of magnitude brighter than those in our sample.  \cite{Aird12} finds
that the distribution of Eddington ratios is independent of stellar
mass, suggesting that the physical mechanism responsible for fueling
AGN is similar at all scales and times.  Our results support this
picture, with no increase in X-ray luminosity per unit stellar mass
from $9.4 \lesssim \log M/M_{\odot} \lesssim 11.6$

Making the plausible assumption that the non-stellar X-ray emission is
dominated by AGN, our study suggests that black holes have higher
accretion rates in star forming galaxies than 
quiescent galaxies, though there is significant scatter in the
relation. This result may seem in contrast to many studies that found a lack of SFR-AGN connection up to
$z\sim3$ \citep{Symeonidis11,CD12, Farrah12,Rosario12,Rosario13}.  We
note that these studies
have not corrected for the mass evolution of their sample, and thus
probe different galaxy populations at different redshifts.
\cite{Mullaney12} find that AGN host galaxies have SSFRs $\sim 20\%$
lower relative to a normal galaxy sample out to $z\sim3$, but do
not probe the lowest X-ray luminosity regimes, selecting only sources
with $L_x >10^{42}$.  This paper explores a different regime than the high-SFR, high accretion
rates of the bright AGNs that are likely associated with mergers
(e.g. \citealt{Santini12}).  Low
luminosity AGNs are most likely fueled by minor mergers and disk
instabilities, and may accrete at low rates for periods which long
outlast any intense periods of induced star formation
(\cite{Schawinski11} and references therein).

While this study probes a large range of galaxy SED types, masses, and
redshifts, we note several limitations of this study.  First, by
stacking galaxies we are observing the average detection within the
stack; unusually luminous galaxies could dominate the stacked
detection. We compare to the C-COSMOS point source catalog
\citep{Puccetti09}, and find that in our stacked galaxy populations, with the
exception of a few high-redshift bins in which one lone X-ray source
is detected, non-detections typically provide $>85 \%$ of the stack
luminosity.  This suggests that luminosities of individual bright AGNs do
not dominate the stacks.  This is perhaps unsurprising, as this study
automatically eliminates the more unusual SEDs of Type I AGNs by
fitting to galaxy spectra when deriving photometric redshifts (a vast
majority of low-luminosity AGNs appear obscured, while those at higher
luminosity are primarily unobscured; see, e.g.,
\citealt{Merloni04,Hasinger08, Treister10}).  Bias from the removal of
Type I AGNs may be most prominent for the high mass galaxy sample, as
these galaxies are more frequently AGN hosts and have a higher
unobscured fraction (e.g. \citealt{Georgak11,Fanidakis12,Hopkins12}).
Galaxies with SEDs that do not resemble any galaxy SED
template (17\% of the sample) average a full-band luminosity of $10^{42.0}$
erg/s, significantly higher than most unnormalized SED types,
suggesting that AGNs may contribute significantly to the luminosities and
SED shapes of these galaxies.

\section{Summary}

We explore the X-ray properties of $\sim 3500$ galaxies at $0.5<z<2.0$
by stacking the galaxies as a function of SED type
(categorized by 32 SED templates in \citealt{Kriek11}), while
controlling for mass and redshift.
We find that $L_X$ is roughly linearly correlated with galaxy mass, and also increases slightly with increasing redshift. After normalizing by galaxy mass and correcting for the redshift correlation, X-ray luminosity decreases with an increase in stellar population ages (as measured by $D(4000)$).

We utilize local XRB luminosity functions to determine the stellar
contribution to the galaxy X-ray luminosities.  We find that there is
a significant luminosity excess for most galaxies, particularly those
with high values of $D(4000)$, suggesting large X-ray contributions
from hot gas and/or black hole accretion.  
 
Consistent with other studies, we find that low luminosity AGNs show at best a mild
preference for galaxies with shallower $4000$\AA~breaks and higher SSFRs (e.g.,
\citealt{Xue10,Georgak11,Rosario11,Hainline12,Kocevski12}). First, there is a
large scatter in X-ray luminosity from non-stellar sources for similar
galaxy types. Second, there is no strong correlation between the X-ray
emission from non-stellar sources and spectral type.  However, we
observe that obscuration (as determined by hardness
ratio) appears to increase with increasing redshift as the photon
index decreases.  This is consistent with the slight increase in X-ray
luminosity per unit stellar mass with redshift, suggesting an increase
in AGN accretion rates in these galaxies.
 We note that we do not probe the highest
black hole accretion rates, as our selection methods bias against galaxies for which the continuum emission is not dominated by stellar light.

While it is difficult to differentiate sources of X-ray emission from
C-COSMOS data alone, a similar analysis of COSMOS radio data (such as
that undertaken by \cite{Carilli08} of Lyman Break Galaxies in the
field), mid-infrared data, and 6-79keV X-ray data from NuSTAR could
provide more direct information about the obscured AGN activity in
these galaxies, helping to differentiate them from heavily
star-forming galaxies that are X-ray hard.  In addition, NIR spectra
of $z \sim 1.5$ galaxies can aid in more
accurate measurements of AGN demographics using emission line ratios
[N{\sc II}]/H$\alpha$ and [O{\sc III}]/H$\beta$.  An analogous
analysis using the deeper Chandra Deep Field South data could also be
performed, helping to constrain the amount of variability within
individual galaxy SED types.

We would like to acknowledge the NMBS and the Chandra COSMOS
teams for their observations and creation of the catalogs.  We would like to thank Ryan Hickox and Jane
Rigby for useful discussions.  T.J. acknowledges support of the
National Science Foundation Graduate Research Fellowship.


\vspace{20mm}
\begin{thebibliography}{XXX}

\bibitem[Aird et al.(2012)]{Aird12} Aird, J., Coil, A.~L., 
Moustakas, J., et al.\ 2012, \apj, 746, 90 

\bibitem[Ballantyne(2008)]{Ballantyne08} Ballantyne, D.~R.\ 2008, 
\apj, 685, 787 

\bibitem[Barmby et al.(2008)]{Barmby08} Barmby, P., Huang, 
J.-S., Ashby, M.~L.~N., et al.\ 2008, \apjs, 177, 431

\bibitem[Bongiorno et al.(2012)]{Bongiorno12} Bongiorno, A., 
Merloni, A., Brusa, M., et al.\ 2012, arXiv:1209.1640 

\bibitem[Bournaud et al.(2011)]{Bournaud11} Bournaud, F., Dekel, 
A., Teyssier, R., et al.\ 2011, \apjl, 741, L33 

\bibitem[Brammer et al.(2008)]{Brammer08} Brammer, G.~B., van 
Dokkum, P.~G., \& Coppi, P.\ 2008, \apj, 686, 1503 

\bibitem[Brammer et al.(2009)]{Brammer09} Brammer, G.~B., 
Whitaker, K.~E., van Dokkum, P.~G., et al.\ 2009, \apjl, 706, L173 

\bibitem[Brammer et al.(2011)]{Brammer11} Brammer, G.~B., 
Whitaker, K.~E., van Dokkum, P.~G., et al.\ 2011, \apj, 739, 24 

\bibitem[Bruzual A.(1983)]{Bruzal83} Bruzual A., G.\ 1983, \apj, 
273, 105 

\bibitem[Bruzual 
\& Charlot(2003)]{BC03} Bruzual, G., \& Charlot, S.\ 2003, \mnras, 344, 1000 

\bibitem[Calzetti et al.(2000)]{Calzetti00} Calzetti, D., Armus, 
L., Bohlin, R.~C., et al.\ 2000, \apj, 533, 682 

\bibitem[Canalizo 
\& Stockton(2001)]{Canalizo01} Canalizo, G., \& Stockton, A.\
2001, \apj, 555, 719

\bibitem[Cano-D{\'{\i}}az et 
al.(2012)]{CD12} Cano-D{\'{\i}}az, M., Maiolino, R.,
Marconi, A., et al.\ 2012, \aap, 537, L8 

\bibitem[Capak et al.(2007)]{Capak07} Capak, P., Aussel, H., 
Ajiki, M., et al.\ 2007, \apjs, 172, 99 

\bibitem[Cardamone et al.(2010)]{Cardamone10} Cardamone, C.~N., van 
Dokkum, P.~G., Urry, C.~M., et al.\ 2010, \apjs, 189, 270

\bibitem[Carilli et al.(2008)]{Carilli08} Carilli, C.~L., Lee, 
N., Capak, P., et al.\ 2008, \apj, 689, 883 

\bibitem[Churazov et al.(2002)]{Churazov02} Churazov, E., Sunyaev, 
R., Forman, W., {\&Bouml}hringer, H.\ 2002, \mnras, 332, 729 

\bibitem[Croton et al.(2006)]{Croton06} Croton, D.~J., Springel, 
V., White, S.~D.~M., et al.\ 2006, \mnras, 365, 11 

\bibitem[David et al.(2006)]{David06} David, L.~P., Jones, C., 
Forman, W., Vargas, I.~M., \& Nulsen, P.\ 2006, \apj, 653, 207 

\bibitem[Davies et al.(2007)]{Davies07} Davies, R.~I., 
M{\"u}ller S{\'a}nchez, F., Genzel, R., et al.\ 2007, \apj, 671, 1388

\bibitem[Diamond-Stanic 
\& Rieke(2012)]{DS12} Diamond-Stanic, A.~M., \& Rieke,
G.~H.\ 2012, \apj, 746, 168 

\bibitem[Elvis et al.(2009)]{Elvis09} Elvis, M., Civano, F., 
Vignali, C., et al.\ 2009, \apjs, 184, 158

\bibitem[Erben et al.(2009)]{Erben09} Erben, T., Hildebrandt, H., Lerchster, M., et al.\ 2009, \aap, 493, 1197 

\bibitem[Fabbiano(1989)]{Fabbiano89} Fabbiano, G.\ 1989,
  \araa, 27, 87 

\bibitem[Faber et al.(2007)]{Faber07} Faber, S.~M., Willmer, 
C.~N.~A., Wolf, C., et al.\ 2007, \apj, 665, 265 

\bibitem[Fanidakis et al.(2012)]{Fanidakis12} Fanidakis, N., Baugh, 
C.~M., Benson, A.~J., et al.\ 2012, \mnras, 419, 2797 

\bibitem[Farrah et al.(2012)]{Farrah12} Farrah, D., Urrutia, T., 
Lacy, M., et al.\ 2012, \apj, 745, 178 

\bibitem[Ferrarese 
\& Merritt(2000)]{Ferrarese00} Ferrarese, L., \& Merritt, D.\
2000, \apjl, 539, L9

\bibitem[Gebhardt et al.(2000)]{Gebhardt00} Gebhardt, K., Bender, 
R., Bower, G., et al.\ 2000, \apjl, 539, L13 

\bibitem[Georgakakis et al.(2011)]{Georgak11} Georgakakis, A., 
Coil, A.~L., Willmer, C.~N.~A., et al.\ 2011, \mnras, 418, 2590 

\bibitem[Georgantopoulos et 
al.(2011)]{Georg11} Georgantopoulos, I., Rovilos, E., \&
Comastri, A.\ 2011, \aap, 526, A46 

\bibitem[Gilfanov(2004)]{Gilfanov04a} Gilfanov, M.\ 2004, \mnras, 
349, 146 

\bibitem[Gilfanov et al.(2004)]{Gilfanov04b} Gilfanov, M., Grimm, 
H.-J., \& Sunyaev, R.\ 2004, \mnras, 347, L57 

\bibitem[Grimm et al.(2003)]{Grimm03} Grimm, H.-J., Gilfanov, 
M., \& Sunyaev, R.\ 2003, \mnras, 339, 793 

\bibitem[Hainline et al.(2012)]{Hainline12} Hainline, K.~N., 
Shapley, A.~E., Greene, J.~E., et al.\ 2012, arXiv:1206.3308 

\bibitem[Hasinger(2008)]{Hasinger08} Hasinger, G.\ 2008, \aap, 490, 905 

\bibitem[Hildebrandt et 
al.(2009)]{Hildebrandt09} Hildebrandt, H., Pielorz, J., Erben,
T., et al.\ 2009, \aap, 498, 725 

\bibitem[Ho(2009)]{Ho09} Ho, L.~C.\ 2009, \apj, 699, 

\bibitem[Hogg(2001)]{Hogg01} Hogg, D.~W.\ 2001, 
arXiv:astro-ph/0105280 

\bibitem[Hopkins(2011a)]{Hopkins11a} Hopkins, P.~F.\ 2011, 
arXiv:1101.4230

\bibitem[Hopkins et al.(2011b)]{Hopkins11b} Hopkins, P.~F., 
Hayward, C.~C., Narayanan, D., \& Hernquist, L.\ 2011, \mnras, 2115 

\bibitem[Hopkins et al.(2012)]{Hopkins12} Hopkins, P.~F., 
Hayward, C.~C., Narayanan, D., \& Hernquist, L.\ 2012, \mnras, 420,
320 

\bibitem[Kauffmann et al.(2003)]{Kauffmann03} Kauffmann, G., 
Heckman, T.~M., Tremonti, C., et al.\ 2003, \mnras, 346, 1055 

\bibitem[Kocevski et al.(2012)]{Kocevski12} Kocevski, D.~D., 
Faber, S.~M., Mozena, M., et al.\ 2012, \apj, 744, 148 

\bibitem[Kriek et al.(2009)]{Kriek09} Kriek, M., van Dokkum, 
P.~G., Labb{\'e}, I., et al.\ 2009, \apj, 700, 221

\bibitem[Kriek et al.(2010)]{Kriek10} Kriek, M., Labb{\'e}, I., 
Conroy, C., et al.\ 2010, \apjl, 722, L64 

\bibitem[Kriek et al.(2011)]{Kriek11} Kriek, M., van Dokkum, 
P.~G., Whitaker, K.~E., et al.\ 2011, \apj, 743, 168 

\bibitem[Labb{\'e} et al.(2005)]{Labbe05} Labb{\'e}, I., Huang, 
J., Franx, M., et al.\ 2005, \apjl, 624, L81 

\bibitem[Laird et al.(2010)]{Laird10} Laird, E.~S., Nandra, K., 
Pope, A., \& Scott, D.\ 2010, \mnras, 401, 2763 

\bibitem[Lehmer et al.(2007)]{Lehmer07} Lehmer, B.~D., Brandt, 
W.~N., Alexander, D.~M., et al.\ 2007, \apj, 657, 681

\bibitem[Lehmer et al.(2008)]{Lehmer08} Lehmer, B.~D., Brandt, 
W.~N., Alexander, D.~M., et al.\ 2008, \apj, 681, 1163 

\bibitem[Madau et al.(1998)]{Madau98} Madau, P., Pozzetti, L., 
\& Dickinson, M.\ 1998, \apj, 498, 106 

\bibitem[Magorrian et al.(1998)]{Magorrian98} Magorrian, J., 
Tremaine, S., Richstone, D., et al.\ 1998, \aj, 115, 2285 

\bibitem[Marchesini et al.(2010)]{Marchesini10} Marchesini, D., 
Whitaker, K.~E., Brammer, G., et al.\ 2010, \apj, 725, 1277

\bibitem[Merloni et al.(2004)]{Merloni04} Merloni, A., Rudnick, 
G., \& Di Matteo, T.\ 2004, \mnras, 354, L37 

\bibitem[Mullaney et al.(2012)]{Mullaney12} Mullaney, J.~R., 
Pannella, M., Daddi, E., et al.\ 2012, \mnras, 419, 95 

\bibitem[Norman et al.(2004)]{Norman04} Norman, C., Ptak, A., 
Hornschemeier, A., et al.\ 2004, \apj, 607, 721 

\bibitem[Pellegrini 
\& Ciotti(1998)]{Pellegrini98} Pellegrini, S., \& Ciotti, L.\
1998, \aap, 333, 433 

\bibitem[Pellegrini(2010)]{Pellegrini10} Pellegrini, S.\ 2010, \apj, 
717, 640 

\bibitem[Persic 
\& Rephaeli(2007)]{PR07} Persic, M., \& Rephaeli, Y.\
2007, \aap, 463, 481

\bibitem[Puccetti et al.(2009)]{Puccetti09} Puccetti, S., Vignali, 
C., Cappelluti, N., et al.\ 2009, \apjs, 185, 586

\bibitem[Rosario et al.(2011)]{Rosario11} Rosario, D.~J., Mozena, 
M., Wuyts, S., et al.\ 2011, arXiv:1110.3816 

\bibitem[Rosario et al.(2012)]{Rosario12} Rosario, D.~J., 
Santini, P., Lutz, D., et al.\ 2012, arXiv:1203.6069 

\bibitem[Rosario et al.(2013a)]{Rosario13} Rosario, D.~J., Mozena, 
M., Wuyts, S., et al.\ 2013, \apj, 763, 59 

\bibitem[Rosario et al.(2013b)]{Rosario13b} Rosario, D.~J., 
Santini, P., Lutz, D., et al.\ 2013, \apj, 771, 63

\bibitem[Salpeter(1955)]{Salpeter55} Salpeter, E.~E.\ 1955, \apj, 
121, 161 

\bibitem[Sarazin et al.(2001)]{Sarazin01} Sarazin, C.~L., Irwin, 
J.~A., \& Bregman, J.~N.\ 2001, \apj, 556, 533

\bibitem[Sanders et al.(2007)]{Sanders07} Sanders, D.~B., 
Salvato, M., Aussel, H., et al.\ 2007, \apjs, 172, 86 

\bibitem[Santini et al.(2012)]{Santini12} Santini, P., Rosario, 
D., Shao, L., et al.\ 2012, arXiv:1201.4394 

\bibitem[Schawinski et al.(2009)]{Schawinski09} Schawinski, K., 
Virani, S., Simmons, B., et al.\ 2009, \apjl, 692, L19 

\bibitem[Schawinski et al.(2010)]{Schawinski10} Schawinski, K., 
Urry, C.~M., Virani, S., et al.\ 2010, \apj, 711, 284 

\bibitem[Schawinski et al.(2011)]{Schawinski11} Schawinski, K., 
Treister, E., Urry, C.~M., et al.\ 2011, \apjl, 727, L31 

\bibitem[Scoville et al.(2007)]{Scoville07} Scoville, N., Aussel, 
H., Brusa, M., et al.\ 2007, \apjs, 172, 1 

\bibitem[Shapley et al.(2001)]{Shapley01} Shapley, A., Fabbiano, 
G., \& Eskridge, P.~B.\ 2001, \apjs, 137, 139

\bibitem[Silverman et al.(2009)]{Silverman09} Silverman, J.~D., 
Lamareille, F., Maier, C., et al.\ 2009, \apj, 696, 396 

\bibitem[Symeonidis et al.(2011)]{Symeonidis11} Symeonidis, M., 
Georgakakis, A., Seymour, N., et al.\ 2011, \mnras, 417, 2239 

\bibitem[Tremmel et al.(2013)]{Tremmel13} Tremmel, M., Fragos, 
T., Lehmer, B.~D., et al.\ 2013, \apj, 766, 19 

\bibitem[Treister et al.(2010)]{Treister10} Treister, E., 
Natarajan, P., Sanders, D.~B., et al.\ 2010, Science, 328, 600 

\bibitem[van de Voort et al.(2011)]{Freeke11a} van de Voort, F., 
Schaye, J., Booth, C.~M., \& Dalla Vecchia, C.\ 2011, \mnras, 415,
2782 

\bibitem[van de Voort et al.(2011)]{Freeke11b} van de Voort, F., 
Schaye, J., Booth, C.~M., Haas, M.~R., 
\& Dalla Vecchia, C.\ 2011, \mnras, 414, 2458 


\bibitem[van Dokkum et al.(2009)]{vD09} van Dokkum, P.~G., 
Labb{\'e}, I., Marchesini, D., et al.\ 2009, \pasp, 121, 2 

\bibitem[van Dokkum et al.(2010)]{vD10} van Dokkum, P.~G., 
Whitaker, K.~E., Brammer, G., et al.\ 2010, \apj, 709, 1018 

\bibitem[van Dokkum et al.(2011)]{vD11} van Dokkum, P.~G., 
Brammer, G., Fumagalli, M., et al.\ 2011, \apjl, 743, L15

\bibitem[Watson et al.(2009)]{Watson09} Watson, C.~R., Kochanek, 
C.~S., Forman, W.~R., et al.\ 2009, \apj, 696, 2206 

\bibitem[Whitaker et al.(2011)]{Whitaker11} Whitaker, K.~E., 
Labb{\'e}, I., van Dokkum, P.~G., et al.\ 2011, \apj, 735, 86 

\bibitem[Wild et al.(2010)]{Wild10} Wild, V., Heckman, T., 
\& Charlot, S.\ 2010, \mnras, 405, 933 

\bibitem[Xue et al.(2010)]{Xue10} Xue, Y.~Q., Brandt, W.~N., 
Luo, B., et al.\ 2010, \apj, 720, 368 

\bibitem[Wake et al.(2011)]{Wake11} Wake, D.~A., Whitaker, 
K.~E., Labb{\'e}, I., et al.\ 2011, \apj, 728, 46 

\bibitem[Whitaker et al.(2010)]{Whitaker10} Whitaker, K.~E., van 
Dokkum, P.~G., Brammer, G., et al.\ 2010, \apj, 719, 1715 

\bibitem[Williams et al.(2009)]{Williams09} Williams, R.~J., 
Quadri, R.~F., Franx, M., van Dokkum, P., 
f\& Labb{\'e}, I.\ 2009, \apj, 691, 1879 

\bibitem[Wuyts et al.(2007)]{Wuyts07} Wuyts, S., Labb{\'e}, I., 
Franx, M., et al.\ 2007, \apj, 655, 51 

\end{thebibliography}
\end{document}